\documentclass[aip,amsmath,amssymb,reprint]{revtex4-1}
\usepackage{float}
\usepackage{ulem}
\usepackage[euler]{textgreek}
\usepackage{mathrsfs}
\usepackage[T1]{fontenc}
\usepackage{bm}
\usepackage{graphicx}
\usepackage{amsmath}
\usepackage[utf8]{inputenc}
\usepackage{textcomp}
\usepackage{upgreek}
\usepackage[dvipsnames]{xcolor}
\usepackage{hyperref} 
\hypersetup{colorlinks,citecolor=blue, filecolor=blue ,linkcolor=blue , urlcolor=blue, pdftex}
\usepackage{sidecap}

\def \FUW{Institute of Experimental Physics, Faculty of Physics, University of Warsaw, ul. Pasteura 5, 02-093 Warsaw, Poland}

\begin{document}

\title[The optical signature of few-layer ReSe$_2$]{The optical signature of few-layer ReSe$_2$}

\author{\L{}. Kipczak} \affiliation{\FUW}
\author{M.~Grzeszczyk} \affiliation{\FUW}
\author{K. Olkowska-Pucko} \affiliation{\FUW}
\author{A. Babi\'nski} \affiliation{\FUW}
\author{M. R. Molas} \email{maciej.molas@fuw.edu.pl} \affiliation{\FUW}

\date{\today}

\begin{abstract}

Optical properties of thin layers of rhenium diselenide (ReSe$_2$) with thickness ranging from mono- (1~ML) to nona-layer (9~MLs) are demonstrated. The photoluminescence (PL) and Raman scattering were measured at low ($T$=5~K) and room ($T$=300~K) temperature, respectively. The PL spectra of  ReSe$_2$ layers display two well-resolved emission lines, which blueshift by about 120~meV when the layer thickness decreases from 9~MLs to a monolayer. A rich structure of the observed low-energy Raman scattering modes can be explained within a linear chain model. The two phonon modes of intralayer vibrations, observed in Raman scattering spectra at about 120~cm$^{-1}$, exhibit very sensitive and opposite evolution as a function of layer thickness. It is shown that their energy difference can serve as a convenient and reliable tool to determine the thickness of ReSe$_2$ flakes in the few-layer limit.
\end{abstract}

\maketitle

%%%% INTRO %%%%
\section{Introduction \label{sec:Intro}}

\begin{figure}[t!]
	\centering
	\includegraphics[width=\linewidth]{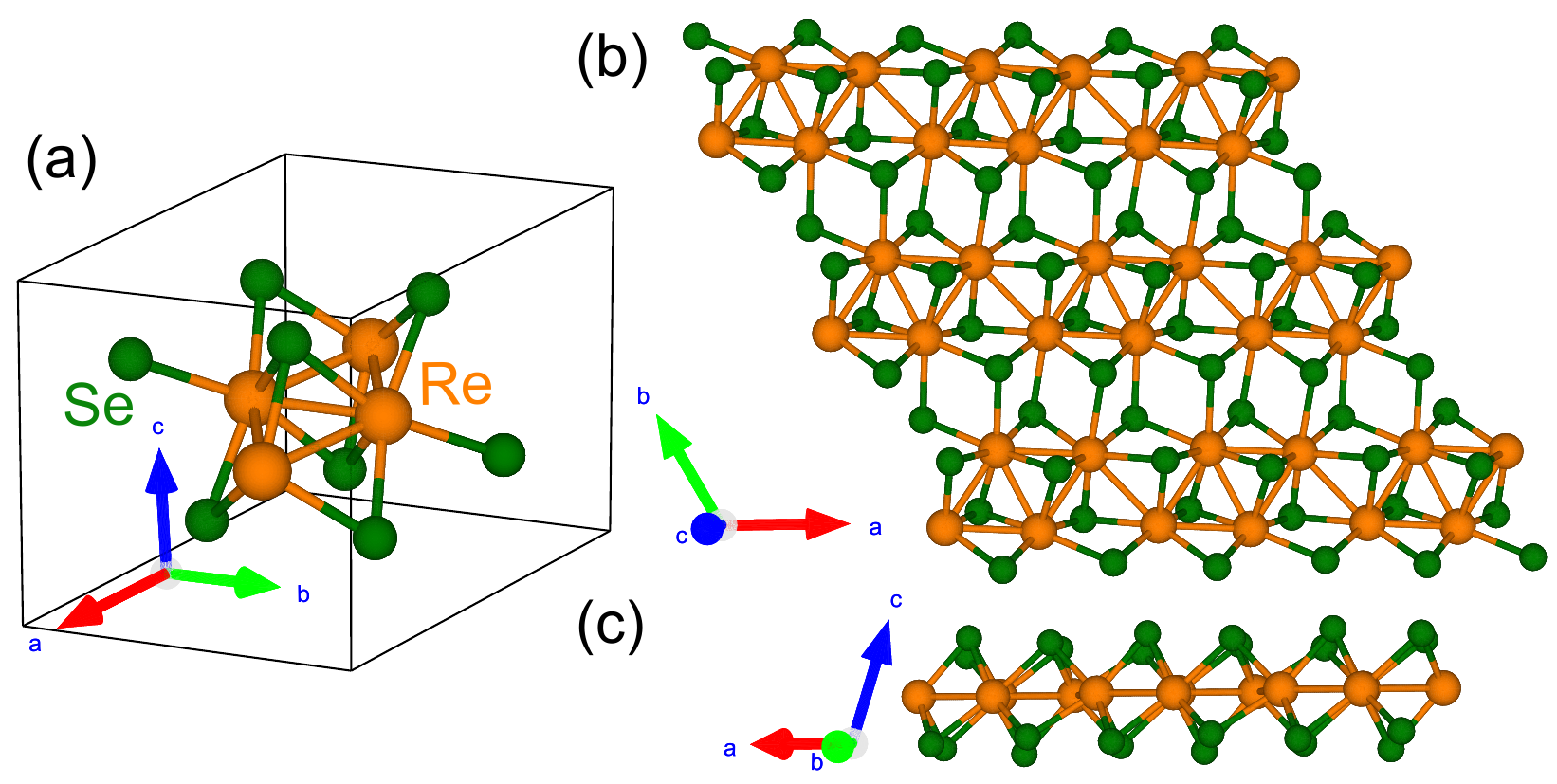}
	\caption{Schematic representation of the 1T$^{'}$-ReSe$_2$ (a) unit cell, (b) top and (c) side view of the monolayer along $c$- and $b$-axis, respectively.}
	\label{fig:structure}
\end{figure}

Semiconducting transition metal dichalcogenides (\mbox{S-TMDs}) such as MoS$_2$, MoSe$_2$, WS$_2$, WSe$_2$ and  MoTe$_2$ invariably attract attention due to their unique electronic structures and resulting optical properties~\cite{Koperski2017,Wang2018}. Recently, rhenium based compounds (ReS$_2$ and ReSe$_2$) with their weak interlayer coupling and in-plane anisotropy, have emerged as intensively investigated members of the S-TMDs family. Like other S-TMDs, ReSe$_2$ crystals are layered materials composed of covalently bonded monolayers stacked together by weak van der Waals interactions. Each  monolayer, made up of three atomic planes (Se-Re-Se), crystallizes in a distorted 1T (1T$^{'}$) phase with triclinic symmetry \cite{Jariwala2017, Lamfers1996}. The schematic representation of the 1T$^{'}$-ReSe$_2$ is demonstrated in Fig.~\ref{fig:structure}. Its anisotropic optical and electronic properties can be utilized to produce field-effect transistors \cite{Nihar2018,ShengxueYang2014,Corbet2016} and photodetectors \cite{Yang2014,Zhang2016}. Moreover, it is reported that thin layer of ReSe$_2$ may be also used to fabricate ReS$_2$/ReSe$_2$ heterojunction functioning as diodes or solar cells\cite{Cho2017}.

The \mbox{S-TMDs} investigation usually starts with the identification of the atomically thin layer. The most common example of a respective tool is based on the Raman scattering spectrum. For instance the energy difference between the in-plane (A$_{1g}$) and out-of-plane (E$^1_{2g}$ ) phonon modes in thin layers of MoS$_2$, which strongly depends on the layer thickness \cite{Lee2010}, enables their correct determination. 

To create a similar tool to distinguish between layers of different thicknesses we have explored optical properties of ReSe$_2$ from mono- (1~ML) to nona-layers (9~MLs). We performed photoluminescence (PL) and Raman scattering (RS) measurements at low ($T$=5~K) and room ($T$=300~K) temperature, respectively. The nature of the band structure of this material remains an open question due to inconsistent results obtained by different groups \cite{gehlmann2017direct, aslan2016linearly,hart2017electronic,aroraReSe2}. Here, we present the low-temperature PL spectra with two well-resolved emission lines, which experience a blueshift of about120~meV with decreasing the number of layers from 9 to 1. This result seems to confirm that the energy gap is direct and increases with the decrease of the layer thickness. The most common and convenient method of characterization of 2D materials is Raman spectroscopy. One of the most accurate means of determining thickness is the analysis of low-frequency interlayer phonons. \cite{chen2015, Lorchat2016, Grzeszczyk2016, Froehlicher2017} Our results present the evolution of low-energy Raman scattering modes with the layer thickness, which we address with a simple linear chain model. However, the ultra-low energy range required to measure these features is not easily accessible. For this reason, we have drawn our attention to the higher energy range, were multiple phonon peaks can be observed in RS spectra. Among them, the energies of two phonon modes of intralayer vibrations, apparent at about 111~cm$^{-1}$ and 120~cm$^{-1}$, exhibit very sensitive thickness dependence. The energy difference between these two peaks decreases from around 12~cm$^{-1}$ for 1~ML to almost 6~cm$^{-1}$ in 7-9~MLs, which may provide a convenient and reliable tool for determining flake thickness in thin layers of ReSe$_2$. We note that this behavior is similar to that observed in ReS$_2$. \cite{Chenet2015}

%%%% METHODS %%%%

\section{Experimental details \label{sec:methods}}

Monolayer and few-layer flakes of ReSe$_2$ were obtained on a Si/(90~nm) SiO$_2$ substrate by polydimethylsiloxane-based exfoliation \cite{Gomez2014} of bulk crystals purchased from HQ Graphene. The flakes of interest were initially identified by visual inspection under an optical microscope. 

The PL and RS measurements were performed using \mbox{$\lambda$=515~nm} (2.41~eV) and \mbox{$\lambda$=633~nm} (1.96~eV) radiation from continuous wave Ar-ion and He-Ne lasers, respectively. The excitation light in those experiments was focused by means of a 100x long-working distance objective with a 0.55 numerical aperture (NA) producing a spot of about 1~$\upmu$m diameter. The signal was collected via the same microscope objective, sent through 0.75~m monochromator, and then detected by using a liquid nitrogen cooled charge-coupled device (CCD) camera. The spectral resolution of our setup in studied energy range of the RS experiment is equal to 0.7~cm$^{-1}$. To detect low-energy RS down to about $\pm$5~cm$^{-1}$ from the laser line, a set of Bragg filters was implemented in both excitation and detection paths. The low-temperature PL measurements were performed with the samples placed on a cold finger in a continuous flow cryostat mounted on $x$-$y$ manual positioners. The excitation power focused on the sample was kept at 200~$\upmu$W during all measurements to avoid local heating.

%%%% RESULTS %%%%

\section{Experimental results \label{results}}

%%%%%%%    PL    %%%%%%%%%

\begin{figure}[t!]
		\centering
		\includegraphics[width=1\linewidth]{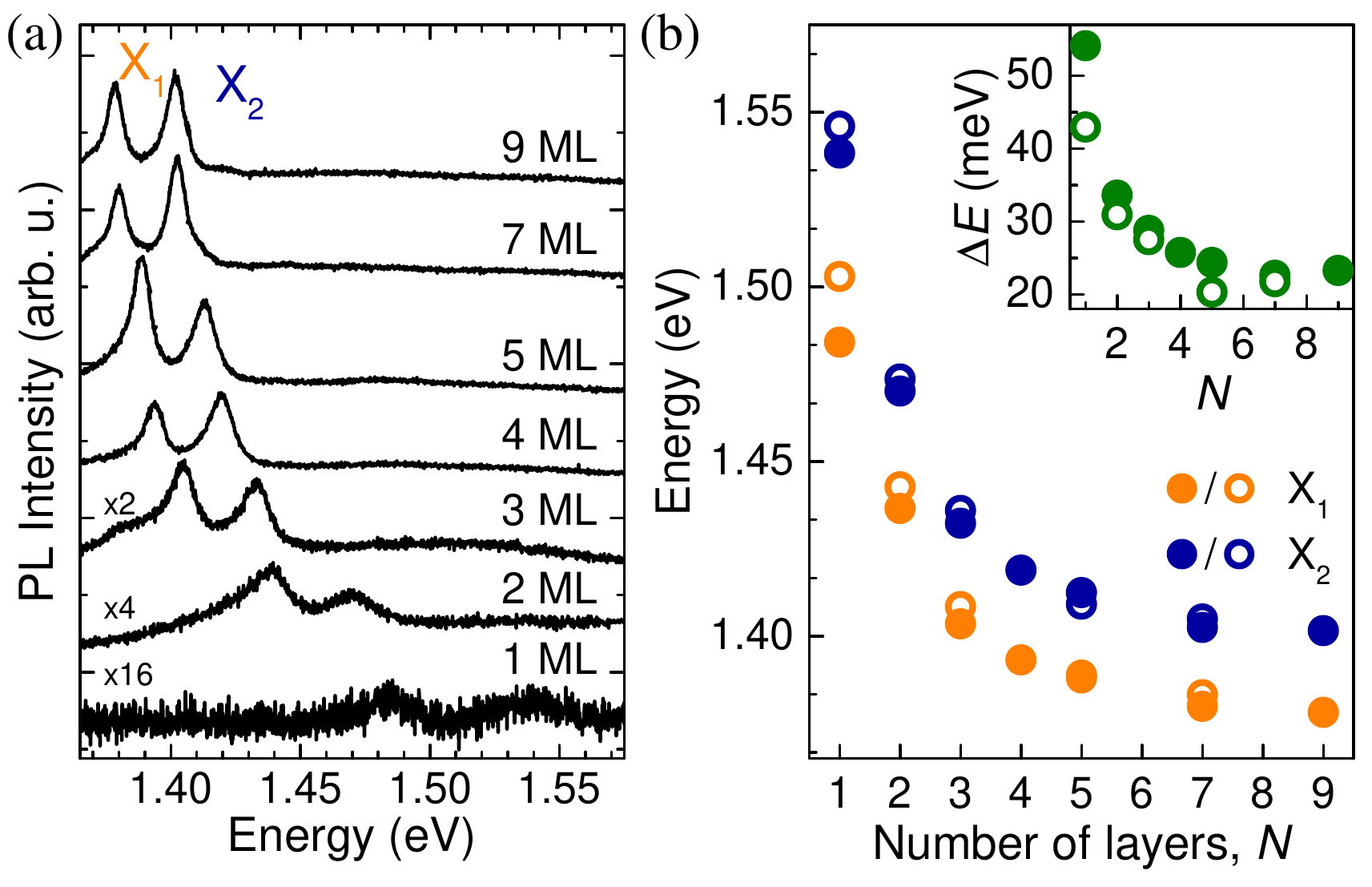}%
		\caption{(a) Photoluminescence spectra of thin ReSe$_2$ flakes of thickness ranging from 1~ML to 9~MLs, measured at $T$=5~K with the use of 2.41~eV laser light excitation. (b) Thickness dependence of the energy of the X$_1$ and X$_2$ peaks. The inset presents the evolution of the energy separation ($\Delta E$) between the X$_1$ and X$_2$ lines. Full points represent experimental data of this work, while the open points denote the results from Ref.~\citenum{aroraReSe2}.}
		\label{fig:pl}
\end{figure}

One of the most recognizable features of molybdenum- and tungsten-based \mbox{S-TMDs} is indirect to direct bandgap crossover in the single-layer limit. However, the \mbox{S-TMD} family also includes materials that do not follow the tendency. One such group are Re-based compounds, namely ReSe$_2$ and ReS$_2$. It was shown that both of these materials behaved like a stack of electronically and vibrationally decoupled monolayers even in bulk form, with their optical properties being almost independent of the number of layers \cite{tongay2014monolayer,Chenet2015,ShengxueYang2014}. In order to examine the ReSe$_2$ thickness evolution of the band structure, we measured the low-temperature PL spectra of flakes consisting of 1~ML up to 9~MLs, presented in Fig.~\ref{fig:pl}(a). The emission spectra comprise two well-resolved peaks referred to as X$_1$ (at lower energy) and X$_2$ (at higher energy). The X$_1$ and X$_2$ lines correspond to the radiative recombination of excitons comprising electrons from the bottom of the conduction band and holes from the top of the valence band. Both of the bands are 2-fold degenerate with each degenerate pair consisting of bands with opposite spins (see Refs~\citenum{aroraReSe2,Ho2019} for details). The excitonic emission peaks shift towards lower energies when the layer thickness is increased, from about 1.51~eV for 1~MLs to $\sim$1.39~eV for 9~MLs (see Fig.~\ref{fig:pl}(b)). Moreover, the energy separation between the X$_1$ and X$_2$ peaks ($\Delta E$) decreases substantially with the increasing number of layers, from about 55~meV for monolayer to almost 20~meV for 9~MLs. The observed evolution of the PL peaks is compared with results of previous low-temperature transmission measurements of thin ReSe$_2$ flakes exfoliated on sapphire substrates demonstrated in Ref.~\citenum{aroraReSe2}. As can be appreciated in Fig.~\ref{fig:pl}(b), the agreement between results obtained from two different types of experiments is quite good. This strongly suggests that the origin of these two investigated transitions is the same in both the emission and absorption measurements. 

Analogous low-temperature PL spectra were measured on thin layers of ReS$_2$ \cite{Jadczak2019ReS2}. Those spectra also comprise two well-defined peaks. Their energies redshift by about 150~meV, when the layer thickness is increased from a monolayer to 15 layers. That redshift is also accompanied by a $\Delta E$ decrease of around 30~meV. The evolution is very similar to that shown in Fig.~\ref{fig:pl}(b), where the corresponding numbers are of the order of 120~meV and 55~meV. These results reveal similar optical and vibrational properties of ReSe$_2$ and ReS$_2$, which crystallize in the same distorted 1T diamond-chain structure.

%%%%%%%%%%%    RS    %%%%%%%%%%%%

The particular 1T$^{'}$ crystalline structure with reduced symmetry, compared to other commonly studied 2H-stacked S-TMDs \cite{Lee2010,Grzeszczyk2016}, affects not only the band structure of ReSe$_2$ but also its vibrational properties. The ReSe$_2$ ML belongs to point group C$_i$, which has only an inversion symmetry \cite{Choi2019, Froehlicher2017}. The unit cell of ReSe$_2$ ML contains 12 atoms, and the irreducible representation can be written as \mbox{$\Gamma$= 18(A$_\textrm{g}$+A$_\textrm{u}$)} \cite{Choi2019}. This gives rise to the equal number (18) of  Raman- and infrared-active modes characterized by the A$_\textrm{g}$ and A$_\textrm{u}$ symmetries, respectively. 

    \begin{figure}[t!]
		\centering
		\includegraphics[width=1\linewidth]{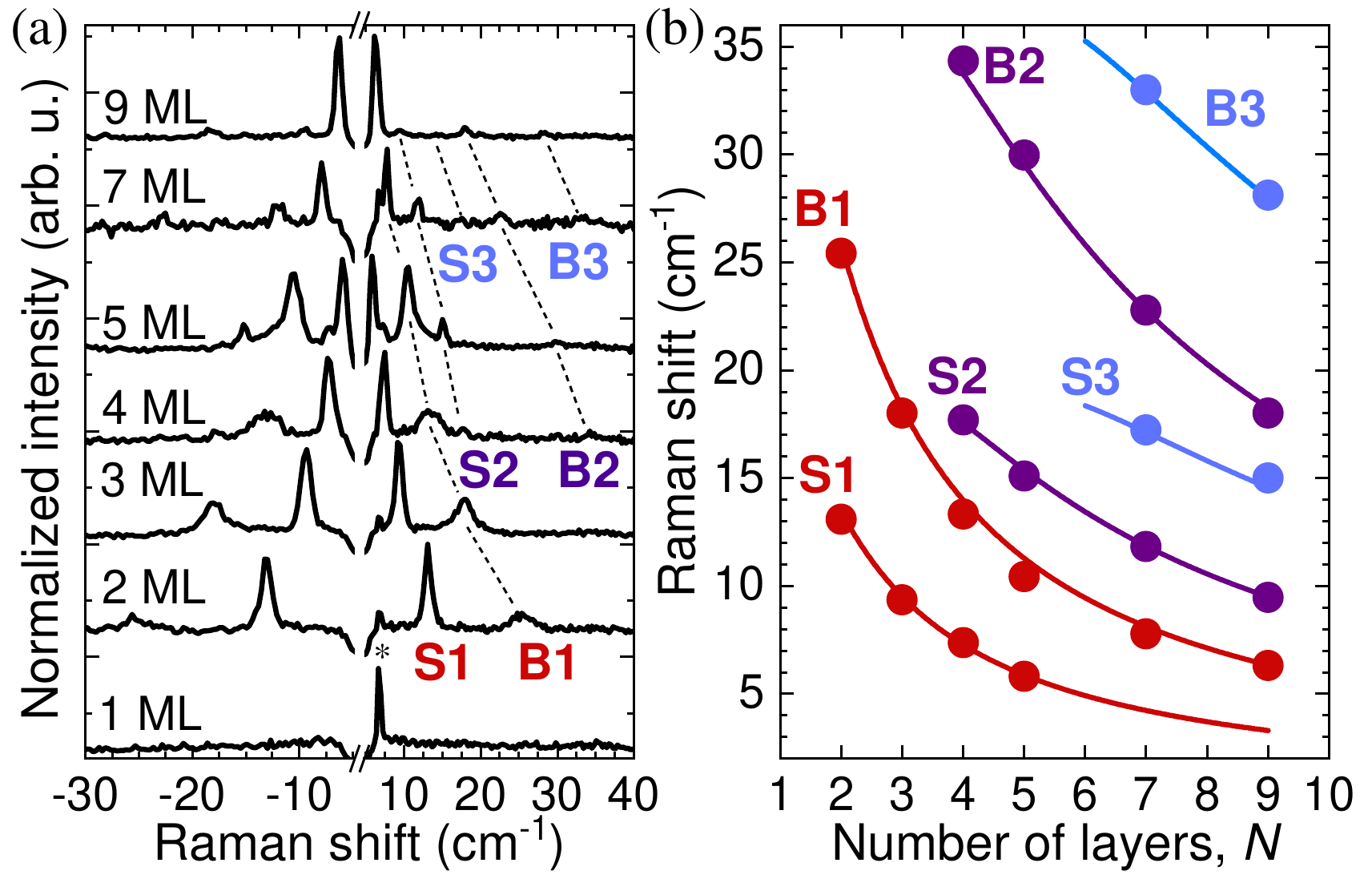}%
		\caption{(a) Low-energy Raman scattering spectra of thin ReSe$_2$ layers with thickness ranging from 1~ML to 9~MLs, measured at $T$=300 K with the use of 1.96~eV laser light excitation. The gray asterisk highlights residual stray light from the exciting laser beam. (b) The evolution of the shear and breathing mode energies with the flake thickness - denoted with circles. The theoretical evolution of these modes is also presented with solid curves.}
		\label{fig:rigid}
	\end{figure}
	
We start our analysis with measurements of low-energy ($<$40~cm$^{-1}$) RS modes, The results of the measurements are presented in Fig.~\ref{fig:rigid}(a). The observed peaks result from rigid interlayer vibrations of the ReSe$_2$ layers. There are two families of the vibrations: shear and breathing modes. They are related to interlayer displacements which are perpendicular and parallel to the crystal $c$-axis respectively. There are no rigid interlayer vibrations in ReSe$_2$ ML, as expected. Two Raman-active modes, $i.e.$ one shear (S1) and one breathing (B1) exist in 2~ML and 3~ML ReSe$_2$. Two shear (S1 and S2) and two breathing (B1 and B2) modes can be observed in 4~ML and 5~ML ReSe$_2$. For thicker flakes ($\geq$5 MLs), an additional peak of both types of vibrations can be appreciated, $i.e.$ B3 and S3. The energy evolution of the shear (S1, S2, and S3) and breathing (B1, B2, and B3) modes with the structure thickness is shown in Fig.~\ref{fig:rigid}(b). This evolution can be well described within a linear chain model, in which a given layer is taken as a single point mass connected to the nearest neighboring layers with springs \cite{lin2019ultralow}. The interlayer interaction can be described by the interlayer force constants $K_i$ ($i$=$s$ or $b$ for shear or breathing modes). The evolution of the interlayer mode energies (expressed in cm$^{-1}$) as a function of the number of layers, $N$, is given by:

\begin{equation}\label{eq:LF}
\omega_{i,\alpha} = \omega_i \sqrt{\left ( 1-\cos\frac{(\alpha-1)\pi}{N}\right )}   ,
\end{equation}

\noindent where $\omega_i$=$\sqrt{{K_i}/{(2\mu\pi^{2}c^{2})}}$. $K_i$ is the respective force constant, $\mu=4\times( 2 m_{\text{Se}} + m_{\text{Re}})$ is the mass per unit cell area ($m_{\text{Se}}=3.4\times10^7$kg/m$^2$ and $m_{\text{Re}}=8.0\times 10^7$kg/m$^2$), $\alpha=2,3,4,\dots,N$ ($\alpha=1$ corresponds to the acoustic mode). The observed shear and breathing modes belong to the lowest-energy branches. The fitted curves are demonstrated in Fig.~\ref{fig:rigid}(b). The simulated curves accurately reproduce the observed low energy peaks. We obtained $\omega_s$=13.4~cm$^{-1}$ and $\omega_b$=25.8~cm$^{-1}$ resulting in the in-plane (out-of-plane) force constant $K_s$ ($K_b$) equal to $18.9\times10^{19}$N/m$^3$ ($69.6\times10^{19}$N/m$^3$), which matches well previously reported data for ReSe$_2$ \cite{Lorchat2016,Froehlicher2017}.

	\begin{figure}[t!]
		\centering
		\includegraphics[width=1\linewidth]{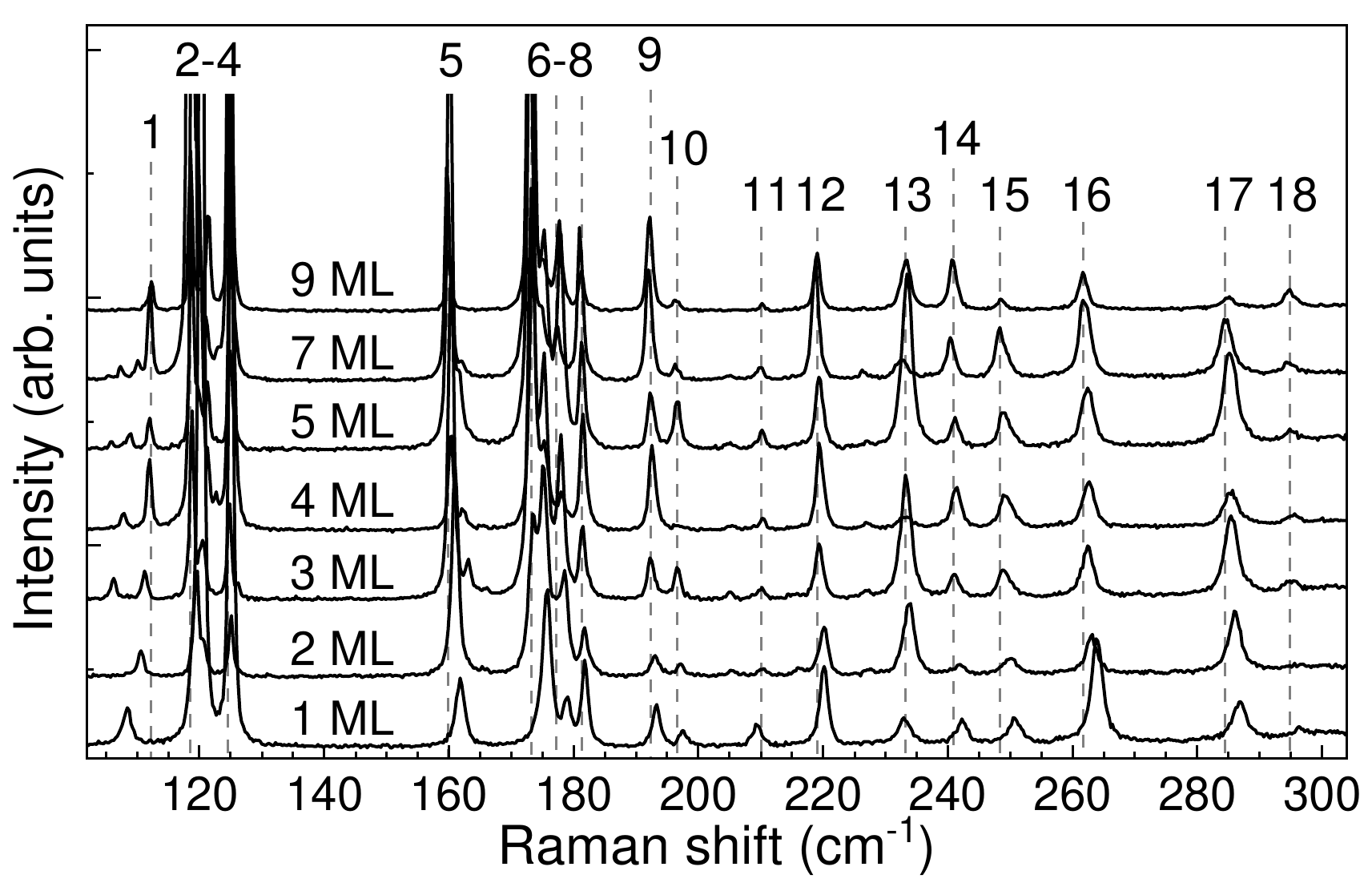}%
		\caption{Raman scattering spectra of thin layers of ReSe$_2$ with thickness ranging from 1~ML to 9~MLs, measured at $T$=300~K with the use of 1.96~eV laser light excitation.}
		\label{fig:raman}
	\end{figure}
	
The RS spectra in higher energy range measured on flakes with thicknesses ranging from 1~ML to 9~MLs ReSe$_2$ are presented in Fig.~\ref{fig:raman}. The in-plane ReSe$_2$ anisotropy results in the extremely rich RS spectra, which are in accordance with published reports \cite{Wolverson2014, Zhao2015, Lorchat2016, Choi2019}. As previously mentioned, there are 18 Raman active peaks, all observed in the frequency range from 100~cm$^{-1}$ to 300~cm$^{-1}$. The considerable sensitivity of some Raman peaks intensities to layer thickness should be noted. It is difficult to conduct a thorough analysis of this behavior without taking into account the orientation of each flake, which is crucial considering the anisotropic nature of the ReSe$_2$ crystal \cite{Choi2019}. To draw substantive conclusions a comprehensive polarization-resolved measurement should be performed, which stays out of the scope of this work.

\begin{figure}[h!]
    	\centering
    	\includegraphics[width=\linewidth]{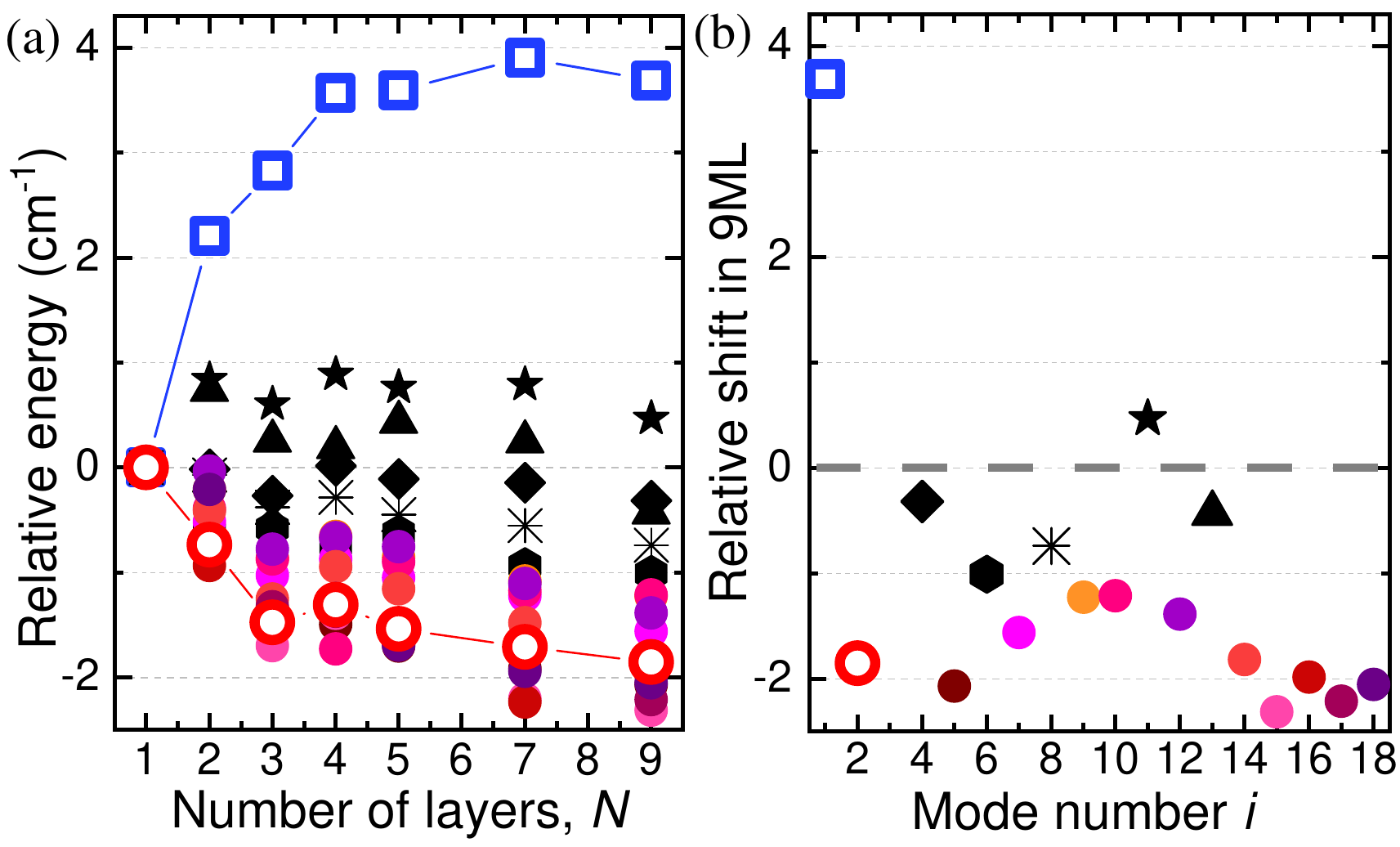}
		\caption{(a) The evolution of the observed Raman scattering peaks energy relative to their energies in 1ML as a function of the flake thickness. (b) The energy difference of a given RS peak between measured on the thickest layer (9~ML) and the thinnest one (1~ML).}
		\label{fig:energie}
\end{figure}

As can be appreciated in Fig.~\ref{fig:energie}(a), the energies of the observed RS modes evolve as a function of the ReSe$_2$ thickness. Basically, three distinct types of dependencies can be noticed. Most of the peaks redshift when the number of layers is increased with the relative energy difference between the 9~ML and 1~ML of 1-2~cm$^{-1}$ (see Fig.~\ref{fig:energie}(b)). There are a few peaks (4, 6, 8, 11, 13), which do not change substantially their energies in the studied range of thicknesses. Finally, peak 1 experiences a significant blueshift of almost 4~cm$^{-1}$ between 1~ML and 9~ML. This distinctive evolutions presented in Fig.~\ref{fig:energie} results from the interplay of interlayer and surface effects \cite{Froehlicher2015,Luo2013}. 

\begin{figure}[t!]
    		\centering
    		\includegraphics[width=1\linewidth]{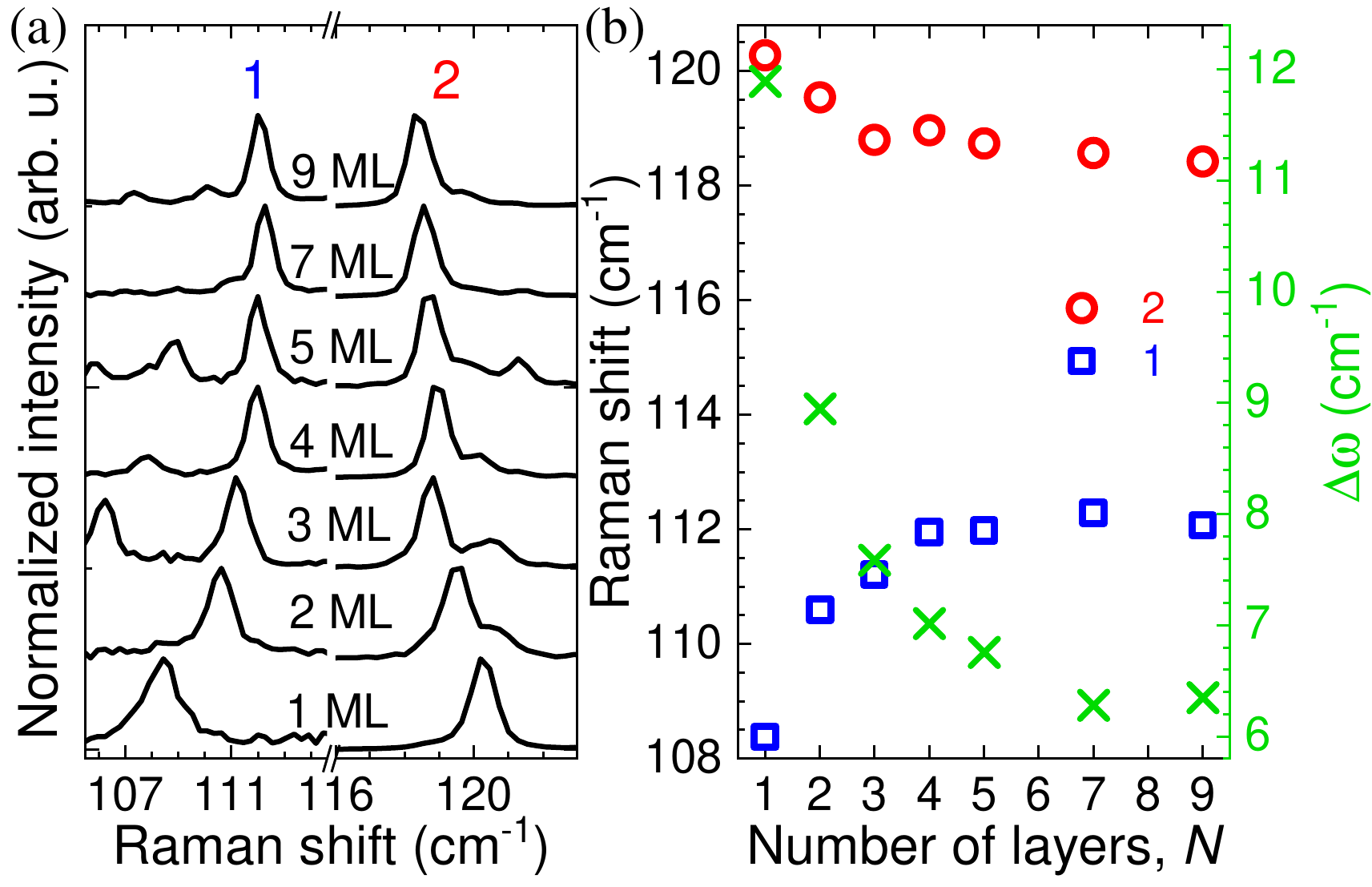}
    		\caption{(a) Raman scattering spectra of thin layers of ReSe$_2$ with thickness ranging from 1~ML to 9~MLs, narrowed to the energy range of the phonon modes, 1 and 2, and measured at $T$=300 K with the use of 1.96~eV laser light excitation. (b) Thickness dependence of energy positions of two phonon modes, 1 and 2, accompanied by their energy difference, $\Delta\omega$.}
    		\label{fig:two_peaks}
\end{figure}

The observed evolution of phonon energies can be very useful to examine the layer's thickness. Let us focus on the specific part of the spectrum, between 100~cm$^{-1}$ and 130~cm$^{-1}$, which is shown in Fig.~\ref{fig:two_peaks}(a). In this range mainly two lines denoted as 1 and 2 can be observed. As already discussed, the phonon mode, labeled 1, blueshifts with an increasing number of layers from 1~ML to 9~MLs, while the peak, labeled 2, displays a redshift. To provide a better illustration of this behavior, we present the extracted energies of both modes for each number of layers in Fig.~\ref{fig:two_peaks}(b). Additionally, the energy difference between peaks 1 and 2 ($\Delta \omega$) is demonstrated in Fig.~\ref{fig:two_peaks}(b), which exhibits a distinct evolution as a function of layer thickness. The $\Delta \omega$ value of almost 12~cm$^{-1}$ for a monolayer is reduced by approximately half in the case of 7-9 layers. This dependency provides a simple and reliable means to identify the sample thickness in a few-layer limit. 

\section{Summary}
In conclusion, we have presented the studies of the optical response (photoluminescence and Raman scattering spectra) of a series of ReSe$_2$ structures with thicknesses varying from a monolayer to nine layers. The low-temperature PL spectra of atomically thin ReSe$_2$ flakes comprise two well-resolved emission lines, which experience a blueshift of about 120~meV when the layer thickness is decreased from 9~MLs to a monolayer limit. The energy of those peaks can be used to determine the layer thickness, but this approach is effectively limited, especially due to the low-temperature conditions of the measurement. The analysis of low-energy interlayer Raman modes allows unambiguous identification of the structure thickness in thin layers of S-TMDs. However, demanding experimental conditions, which enable to approach the excitation laser line as close as 5~cm$^{-1}$, hardly provide a simple characterization tool of S-TMDs. The most convenient way to determine the flake thickness in thin layers of ReSe$_2$ relies on the energy difference between two phonon modes of intralayer vibrations observed at 120~cm$^{-1}$. The energy difference between the peaks changes from 12~cm$^{-1}$ to almost 6~cm$^{-1}$ between 1~ML and 9~MLs. This observation seems to be applicable for other Re-based compounds, $i.e.$ ReS$_2$.

\section*{Acknowledgements}
The work has been supported by the the National Science Centre, Poland (grants no. 2017/27/B/ST3/00205, 2017/27/N/ST3/01612 and 2018/31/B/ST3/02111).

\section*{Data availability}
The data that support the findings of this study are available from the corresponding author upon reasonable request.\\

The following article has been accepted by Journal of Applied Physics. After it is published, it will be found at \url{https://dx.doi.org/10.1063/5.0015289}.

\section*{References}
\bibliography{biblioReSe2}
	
\end{document}